\documentclass[runningheads]{llncs}
\usepackage[T1]{fontenc}
\usepackage{graphicx}
\usepackage{algorithm}
\usepackage[noend]{algpseudocode}
\usepackage{enumitem} 
\usepackage{booktabs}
\usepackage{array}
\usepackage{multirow}
\usepackage{makecell}
\usepackage{bbding}
\usepackage{amsmath}
\usepackage{tabularx}
\usepackage[misc]{ifsym}
\usepackage{amssymb}
\usepackage{booktabs}
\usepackage{multirow}

\begin{document}

\newcommand{\ourmethod}[1]{\texttt{GUSD}}
\title{Unveiling the Hidden:\\
Movie Genre and User Bias in Spoiler Detection}

\author{Haokai Zhang \inst{1} \thanks{H. Zhang and S. Zhang contributed equally to this work.} \and
Shengtao Zhang\inst{1} \and
Zijian Cai\inst{2} \and
Heng Wang\inst{3} \and
Ruixuan Zhu\inst{1} \and
Zinan Zeng\inst{1} \and
Minnan Luo\inst{3,4}(\Letter)
}

\authorrunning{H.Zhang et al.}

\institute{Institute of Artificial Intelligence and Robotics, Xi’an Jiaotong University,
Xi’an 710049, China \\ \email{\{zhanghaokai,zhangst,1760865856,2194214554\}@stu.xjtu.edu.cn}
\and
Institute of Automation, Chinese Academy of Sciences, Beijing, 100190, China \\ \email{caizijian2024@ia.ac.cn}
\and
School of Computer Science and Technology, Xi’an Jiaotong University,
Xi’an 710049, China \\ \email{wh2213210554@stu.xjtu.edu.cn}
\and
Shaanxi Province Key Laboratory of Big Data Knowledge Engineering,
Xi’an Jiaotong University, Xi’an 710049, China \\
\email{minnluo@xjtu.edu.cn}
}

\toctitle{Unveiling the Hidden: Movie Genre and User Bias in Spoiler Detection}
\tocauthor{Haokai~Zhang,Shengtao~Zhang,Zijian~Cai,Heng~Wang,Ruixuan~Zhu,Zinan~Zeng,Minnan~Luo}

\maketitle              
\begin{abstract}
Spoilers in movie reviews are important on platforms like IMDb and Rotten Tomatoes, offering benefits and drawbacks. They can guide some viewers' choices but also affect those who prefer no plot details in advance, making effective spoiler detection essential. Existing spoiler detection methods mainly analyze review text, often overlooking the impact of movie genres and user bias, limiting their effectiveness. To address this, we analyze movie review data, finding genre-specific variations in spoiler rates and identifying that certain users are more likely to post spoilers. Based on these findings, we introduce a new spoiler detection framework called $\textbf{\ourmethod{}}$ (\textbf{G}enre-aware and \textbf{U}ser-specific \textbf{S}poiler \textbf{D}etection), which incorporates genre-specific data and user behavior bias. User bias is calculated through dynamic graph modeling of review history. Additionally, the R2GFormer module combines RetGAT (Retentive Graph Attention Network) for graph information and GenreFormer for genre-specific aggregation. The GMoE (Genre-Aware Mixture of Experts) model further assigns reviews to specialized experts based on genre. Extensive testing on benchmark datasets shows that GUSD achieves state-of-the-art results. This approach advances spoiler detection by addressing genre and user-specific patterns, enhancing user experience on movie review platforms.  Our source code is available at \url{https://github.com/AI-explorer-123/GUSD}

\end{abstract}
\section{Introduction}
\label{introduction}

Spoilers in movie reviews have become an important component of the movie-viewing experience on popular platforms like IMDb and Rotten Tomatoes~\cite{cao2019unifying}. For those who hope to learn the plot of the movie in advance to judge whether they like it or not, the spoilers are helping, while for those who prefer to experience a movie without prior knowledge, the spoilers can severely diminish the enjoyment by revealing crucial plot points, undermining suspense, and eliciting negative emotions among viewers~\cite{loewenstein1994psychology}. Thus, effective spoiler detection methods are crucial for maintaining a positive user experience.

\begin{figure}
    \centering
    \includegraphics[width=0.7\linewidth]{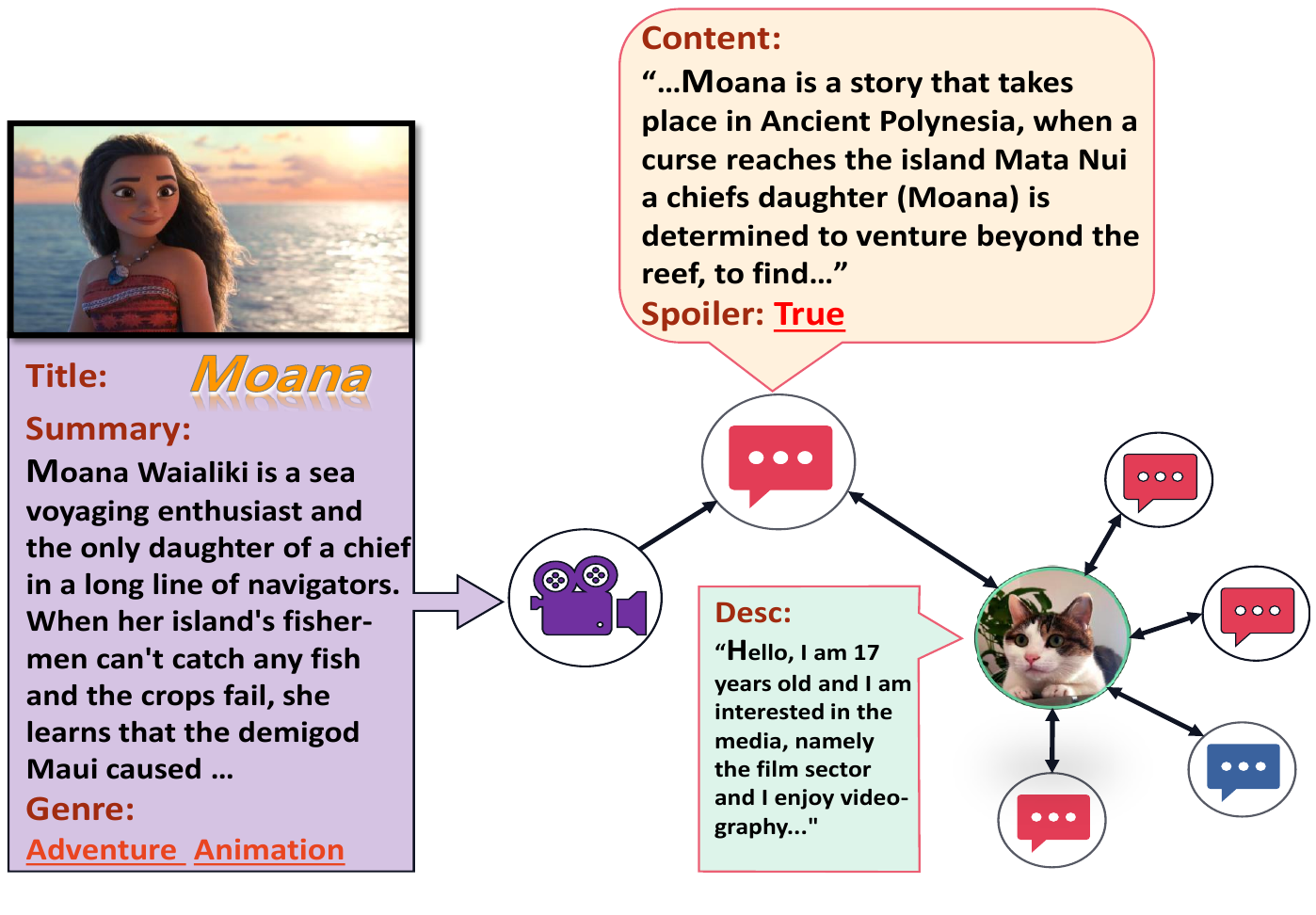}
    \caption{An illustrative example of the data used in our spoiler detection study. The image shows a review of the movie \emph{Moana}, including the movie's genres (Adventure, Animation), summary, the review's content, and user-specific details. All reviews from the user are color-coded: blue indicates non-spoiler content, while red indicates spoiler content. }
    \label{Intro}
\end{figure}
\begin{figure}[tbp]
    \centering
    \includegraphics[width=0.8\linewidth]{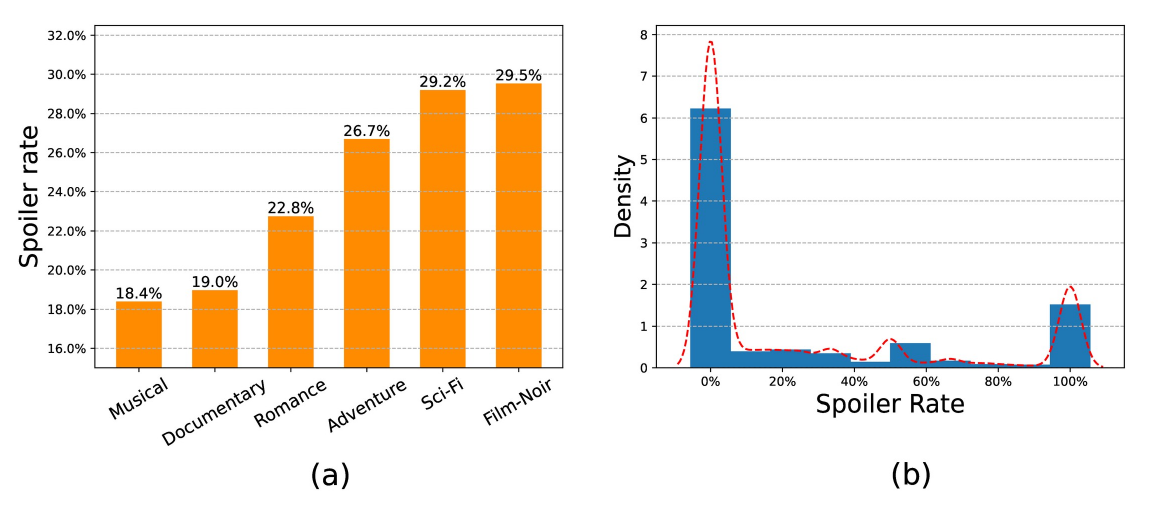}
    \caption{(a) Spoiler rate across different movie genres (partial) in LCS dataset. (b) Kernel density estimation plot and distribution histogram of spoiler across different users.}
    \label{dataset-statistics}
\end{figure}

Existing spoiler detection methods primarily focus on the textual content of reviews. For example, DNSD~\cite{chang2018deep} integrates review sentences and movie genres, while SpoilerNet~\cite{wan2019fine} utilizes a Hierarchical Attention Network and incorporates the item-specificity information. More recent approaches, such as MVSD~\cite{wang2023detecting}, incorporate advanced techniques like syntax-aware graph neural networks and external movie knowledge to improve detection performance. Nevertheless, these methods still exhibit notable limitations. Solely relying on textual content proves insufficient for robust spoiler detection~\cite{wang2023detecting}. Moreover, spoilers are often genre-specific, with varying characteristics depending on the movie's genre — for instance, suspense films focus on plot details, whereas action films emphasize fight scenes. As a result, two significant challenges in spoiler detection remain unaddressed:

\begin{itemize}
    \item \textbf{Diverse Genres}. Previous works have largely ignored the impact of movie genres on the spoiler rate. Our analysis of the dataset indicates substantial differences in spoiler rate across genres, with specific categories defined according to IMDb standards. As shown in Figure \ref{dataset-statistics}(a), movies that heavily rely on plot twists and suspense, such as Film-Noir and Adventure, are more prone to having spoilers in reviews compared to genres like Musical or Documentary. This is understandable since plot-driven movies tend to have more critical plot points that can be spoiled. This variation in spoiler rate demonstrates the importance of considering genre-specific characteristics when developing spoiler detection models. By incorporating genre information, we could better capture these differences and improve the performance of spoiler detection.
    \item \textbf{User-specific Behavior Bias}. User behavior varies significantly, with some users being more prone to posting spoilers than others. Our statistical analysis shows a clear trend where certain users tend to post spoiler reviews more frequently. As illustrated in Figure \ref{dataset-statistics}(b), the graph of user spoiler rate distribution shows that a large proportion of users post very few spoilers, while a smaller, yet significant, group of users frequently post spoilers. This distribution indicates a noticeable user bias, highlighting that certain users are more likely to post spoilers than others. Leveraging these user-specific behavior bias can improve the detection performance by allowing models to adapt to user behavior bias.
\end{itemize}

To address the challenges of genre-specific spoiler tendencies and user bias in spoiler detection, we propose a comprehensive framework named $\textbf{\ourmethod{}}$ (\textbf{G}enre-aware and \textbf{U}ser-specific \textbf{S}poiler \textbf{D}etection). This framework integrates genre information, user behavior bias, and global perception GNN. Our method begins with preprocessing movie, user, and review data. Then user bias is captured from review history through dynamic graph modeling. After that, the core component, R2GFormer (RetGAT and GenreFormer), which consists of RetGAT (Retentive Graph Attention Network) and GenreFormer, processes the graph information. RetGAT aggregates all the features globally, and GenreFormer enhances representation by integrating genre features, allowing for comprehensive genre-specific and cross-genre interactions. And an Aggregator combines these features, and then the GMoE (Genre-Aware Mixture of Experts) model assigns reviews to different experts based on their corresponding movie genres, improving traditional MoE model performance. Finally, a classifier performs spoiler detection using the aggregated features.

Extensive experiments show that \ourmethod{} achieves state-of-the-art performance. We also conduct robustness studies, ablation studies, and specific experiments on GMoE and user bias to validate our proposed modules' effectiveness.

\textbf{Our main contributions are summarized as follows:}
\begin{itemize}
    \item We are the first to model the complex interactions between genre-specific information and long-term user review behaviors for spoiler detection, providing a nuanced approach by understanding genre-specific spoiler characteristics and leveraging user behavior bias.
    \item We propose the \ourmethod{} framework, a novel spoiler detection system that integrates several key components: the GenreFormer to capture genre-specific spoiler tendencies, the GMoE model to dynamically assign reviews based on genres, and dynamic graph modeling to capture user bias. This cohesive integration enhances overall accuracy and robustness.
    \item Our method \ourmethod{} achieves state-of-the-art performance in spoiler detection. Extensive experiments on two benchmark datasets demonstrate its robustness and effectiveness, showing superior performance across various conditions.

\end{itemize}

\section{Related Work}

\subsection{Spoiler Detection}
The goal of automatic spoiler detection is to identify spoilers in reviews from domains like television~\cite{boyd2013spoiler}, books~\cite{wan2019fine}, and movies~\cite{boyd2013spoiler}. Existing approaches to spoiler detection can be broadly classified into three categories: keyword matching methods, machine learning techniques, and deep learning models.

\subsubsection{Keyword matching methods.} These approaches rely on a set of predefined keywords to identify spoilers. Examples include keywords related to sports teams or events~\cite{nakamura2007temporal}, or actors' names~\cite{golbeck2012twitter}. Although useful in specific scenarios, this method requires manual keyword definition and lacks generalizability across different application contexts.
    
\subsubsection{Machine learning techniques.} These methods often involve topic modeling or support vector machines using handcrafted features. For example, Guo et al.~\cite{guo2010finding} applied a bag-of-words representation combined with an LDA-based model for spoiler detection. Jeon et al.~\cite{jeon2013don} developed an SVM classifier incorporating four extracted features, while Boyd et al.~\cite{boyd2013spoiler} utilized lexical features and meta-data of review subjects (e.g., movies and books) in an SVM model.
    
\subsubsection{Deep learning models.} These models mainly leverage NLP techniques, employing RNNs, LSTMs, Transformer, and language models to process review texts and movie information through end-to-end training. Bao et al.~\cite{bao2021spoiler} utilized LSTMs, BERT, and RoBERTa for sentence-level spoiler detection. DNSD~\cite{chang2018deep} focused on incorporating external genre information using GRU and CNN. SpoilerNet~\cite{wan2019fine} introduced item-specificity and bias with bi-RNN enhanced by GRU. SDGNN~\cite{chang2021killing} leveraged dependency relations between context words in sentences with graph neural networks to capture semantics.

While some existing methods incorporate genre information and user bias ~\cite{chang2018deep,wroblewska2021spoiler}, they often rely on quite simple techniques such as concatenating or adding these additional features to the initial review features. Such approaches lack the sophistication needed for more effective and intricate modeling of genre features and user biases. 

\subsection{Mixture of Experts}

The Mixture of Experts (MoE) approach, grounded in the Divide-and-Conquer principle, segments an input sample into sub-tasks and trains specialized experts for each sub-task. This method is extensively utilized in NLP to boost model capacity~\cite{shazeer2017outrageously} and enhance reasoning capabilities~\cite{madaan2021think}. Shazeer et al.~\cite{shazeer2017outrageously} introduced a sparsely-gated Mixture-of-Experts layer, enabling conditional computing in large language models. 
 Fedus et al.~\cite{fedus2022switch} developed simplified routing algorithms for MoE to enhance training stability and reduce computational costs. 
Furthermore, Soft-MoE~\cite{puigcerver2023sparse} was introduced to mitigate issues like training instability and token dropping inherent in traditional MoE approaches.

Despite these advancements, traditional MoE methods assign tokens dynamically, which can cause incorrect token assignment, particularly when the dataset contains explicit category information such as movie genres and their associated reviews. 

\begin{figure}[t]
    \centering  \includegraphics[width=\linewidth]{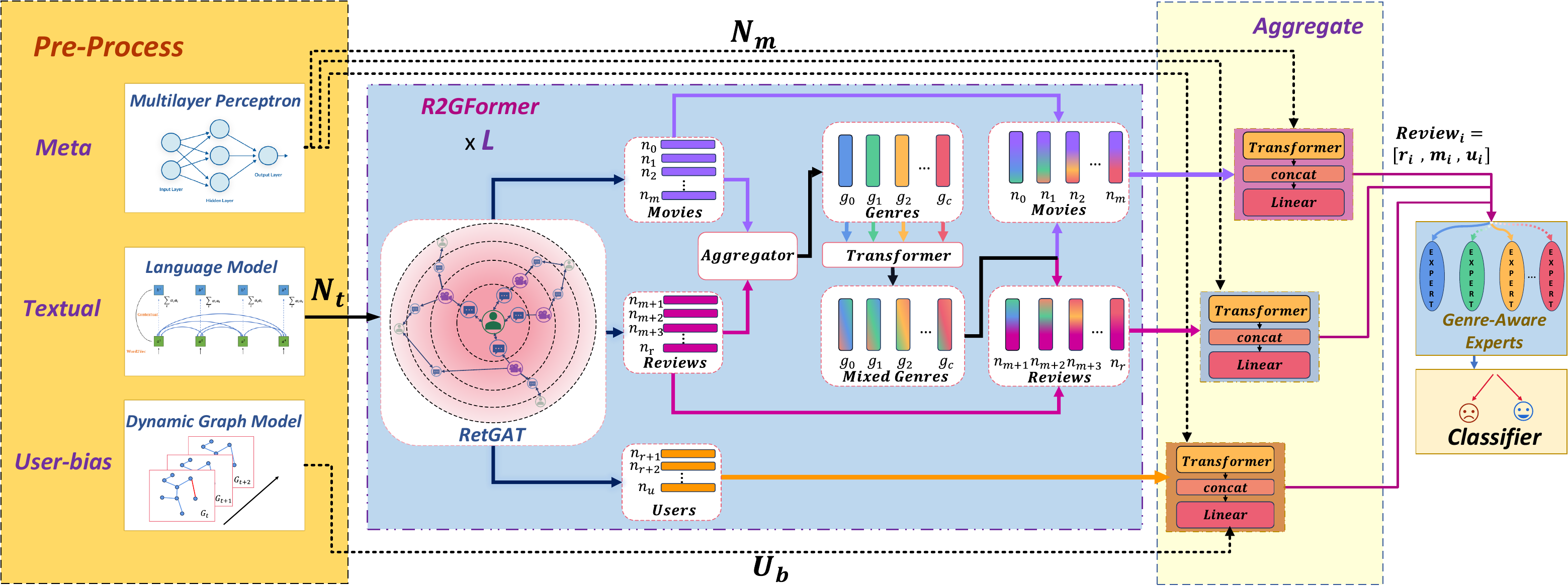}
    \caption{Overview of our proposed \ourmethod{} framework, which integrates genre-specific information, user behavior bias, and global perceptive RetGAT for spoiler detection. It preprocesses movie, user, and review data with MLP and language models, and captures user bias via dynamic graph modeling. Then the data is processed by the R2GFormer component. An Aggregator merges these features, and then GMoE assigns reviews to experts based on genres. Finally, a classifier detects spoilers using the aggregated features.}
    \label{Overview}
\end{figure}

\section{Methodology}

Figure \ref{Overview} shows the architecture of our proposed \ourmethod{} framework. This framework integrates genre-specific information, user behavior bias, and global receptive RetGAT for spoiler detection. Specifically, movie, user, and review data are first to be preprocessed, while user bias is extracted from review history using dynamic graph modeling. The R2GFormer component, consisting of RetGAT and GenreFormer, then processes graph features. RetGAT aggregates graph data, while GenreFormer handles genre-specific data. An Aggregator aggregates these features. The GMoE model assigns reviews to experts based on their related movie genres. Finally, a classifier utilizes the aggregated features to detect spoilers.

\subsection{Data preprocessing}

\subsubsection{Meta information.}

For review nodes, user nodes, and movie nodes, each type possesses metadata (details can be found in the supplementary material). After collecting the metadata for all three types of nodes, we pad them to the same length. A two-layer MLP is then employed as the meta encoder, producing the meta embeddings $N_m$.

\subsubsection{Textual information.}
The textual content is fundamental for effective spoiler detection. To generate high-quality embeddings, we leverage an LM as our text encoder. Specifically, we augment the initial textual content with the textual descriptions of the node's metadata. This augmentation enriches the embeddings by providing additional contextual information about the node. Subsequently, we employ the LM to encode the nodes' textual information. The encoded embeddings are then transformed using a single-layer MLP, producing refined embeddings $N_t$.

\subsubsection{User bias acquisition using dynamic-graph pre-training.}

To better capture the dynamic information attributes of users, we adopt the dynamic graph to handle users' review history flexibly. Thus, we need to convert the static dataset into a dynamic format. We utilize the given connections between different nodes and the time information of the reviews to form the dynamic event stream. The details about the formation of the graph will be displayed in Section \ref{sec:3.2}. Considering the absence of additional information about the dynamic edges, we simply initialize the edge features as zero vectors. Then, we employ the robust DyGFormer~\cite{yu2023towards} as our dynamic graph encoder to capture the interactions among various nodes and utilize Link Prediction as the downstream task. We specifically obtain the features of user nodes as user bias $U_b$ from the dynamic graph encoder.

\subsection{R2GFormer} \label{sec:3.2}

After acquiring the initial meta and textual features of the nodes, we use the textual features $N_t$, which include rich information about the nodes, as the initial embeddings $N_g^{(0)}$ for \textbf{R2GFormer}. This graph encoder not only models the complex relations and interactions between users, reviews, and movies but also handles features from different genres of the entire graph and captures remote dependencies. First, we will introduce how we construct the whole graph. Then, we decompose an R2GFormer layer into two parts: RetGAT and Genreformer, which will be introduced respectively.

\subsubsection{Graph Construction.}
We first construct a directed graph consisting of three types of nodes: \{ \textit{User}, \textit{Review}, \textit{Movie} \} and the following three types of edges:

\underline{E1: \textit{Movie-Review}} We connect a review node to a movie node if the review is about the movie, but not vice versa. This setup allows movie information to influence the review while ensuring that the review information does not affect the movie.

\underline{E2: \textit{Review-User}} We connect a review node to a user node if the review is posted by the user.

\underline{E3: \textit{User-Review}} We connect a user node to a review node if the user posts the review.

\subsubsection{RetGAT.}

Inspired by the work of~\cite{sun2023retentive,fan2024rmt,nikolentzos2020k}, we propose RetGAT, which extends the RetNet framework by integrating a global perception into GAT, incorporating explicit exponential decay for nodes within $k$ hops and truncation for nodes beyond $k$ hops. This method ensures a broad receptive field while balancing performance and computational complexity by dynamically adjusting the influence of nodes based on their distance, with closer nodes having a higher impact on the final node features.

To achieve this global receptive field, we utilize $k$ parallel GAT layers to separately aggregate information from $k$-hop neighbors. For each layer, node features are aggregated from different $k$-hop neighborhoods, applying decay factors to control the influence of information from various hops. Beyond $k$ hops, the influence of nodes is truncated to maintain computational efficiency and focus on relevant information within the $k$-hop range. Note that we correct the algorithm that previous work~\cite{atwood2016diffusion,wang2020multi} used to compute $k$-hop neighbors, and the details are available in the supplementary material.

The decay factor \(\delta_{h}\) for hop $h$ ($h \leq k$) is defined as:
\begin{equation}
\begin{aligned}
    \delta_{h} &= \exp(-\alpha \cdot h),
\end{aligned}
\end{equation}
where \(\alpha\) is a hyperparameter controlling the intensity of exponential decay. For the \(l\)-th layer, the contribution of the \(h\)-th hop neighbors is computed as:
\begin{equation}
\begin{aligned}
    \text{${N_{g,h}^{(l)}}$} &= \delta_{h} \cdot \text{GAT}^{(l)}_h(\mathcal{A}_h, \text{${N_g^{(l-1)}}$}),
\end{aligned}
\end{equation}
where \(\mathcal{A}_h\) represents the adjacency matrix at the \(h\)-th hop, and \(\text{${N_g}^{(l-1)}$}\) denotes the input node features of the $(l-1)$-th layer.

The final node features \(\text{${N_{g}}^{(l)}$}\) for layer \(l\) are aggregated from all the $k$-hop contributions using an Aggregator function:
\begin{align}
    \text{${N_g^{(l)}}$} &= \text{AGGREGATOR}_r^{(l)}(\text{${N_{g,1}^{(l)}}$}, \text{${N_{g,2}^{(l)}}$}, \ldots, \text{${N_{g,h}^{(l)}}$}, \ldots, \text{${N_{g,k}^{(l)}}$}).
\end{align}
where \(\text{AGGREGATOR}_r^{(l)}\) is the aggregator for RetGAT at the \textit{l}-th layer, which can be summation, concatenation, or a TransformerEncoder (TRM).

\subsubsection{Genreformer.}
After message passing among different $k$-hop neighbors, Genreformer aims to pay attention to the entire graph to extract more comprehensive features of different genres. 
First, the global genre-specific representation is obtained by aggregating the features of all the review and movie nodes belonging to the same genre in the $l$-th layer, which is as follows:

\begin{equation}
    g_{j}^{(l)} = \text{AGGREGATOR}^{(l)}_{g}(\{n_{g,i}^{(l)} \mid i \in \mathcal{N}_j\}),
\end{equation}
where $n_{g,i}^{(l)}$ is the feature of node $i$ in the $l$-th layer, $\mathcal{N}_j$ denotes the set of nodes (both review and movie nodes) belonging to the $j$-th genre, and $\text{AGGREGATOR}^{(l)}_{g}$ is the aggregator of Genreformer in the $l$-th layer, which can be summation, concatenation, or a $\text{TRM}$.

Next, we use a $\text{TRM}$ to facilitate inter-genre information interaction, allowing each genre to acquire information from similar genres to further enrich its features:

\begin{equation}
    [g_{1}^{(l)} g_{2}^{(l)} \cdots g_{j}^{(l)} \cdots g_{c}^{(l)}] = \text{TRM}([g_{1}^{(l)} g_{2}^{(l)} \cdots g_{j}^{(l)} \cdots g_{c}^{(l)}]),
\end{equation}
where $c$ is the number of genres.

After interaction among genres, the genre features are fused with the movies and reviews nodes features. Then we incorporate the nodes with their genre features. Since some movies and reviews cover more than one genre, we take the average of the genre features a node covers. Subsequently, we concatenate the review or movie feature and its genre feature, then use an MLP to project it into the desired feature space, $i.e.$,

\begin{align}
    z_i^{(l)} &= \text{MEAN}(\{g_{j}^{(l)} \mid j \in G_{i}\}), \\
    n_{g,i}^{(l)} &= \text{\text{MLP}}([n_{g,i}^{(l)} \parallel z_i^{(l)}]).
\end{align}
where $G_{i}$ denotes the set of genres to which node $i$ belongs, $n_{g,i}^{(l)}$ denotes the feature of node $i$ in the $l$-th layer, $z_i^{(l)}$ is the aggregated genre feature for node $i$ in the $l$-th layer, and $\parallel$ means concatenation. The MLP projects the concatenated feature into the desired feature space.

\subsubsection{Overall interaction.}
One layer of our proposed R2GFormer layer, however, cannot enable the information interaction between all information sources. In order to further facilitate the interaction among the nodes, we employ $L\times$ R2GFormer layers for node representation learning. The representation of the nodes is updated after each layer, incorporating information from different sources. This process can be formulated as follows:

\begin{equation}
    N_{g}^{(l)} = \text{R2GFormer}(N_{g}^{(l-1)}, \mathcal{E}, G).
\end{equation}
where $\mathcal{E}$ denotes all the edges of the sampling graph, $G$ denotes the genres of every movie and review node. After $L\times$ R2GFormer layers, we obtain the final graph representation $N_{g}^{(L)}$ of all nodes.

\subsection{Multimodal Fusion}

Through the data above processed by the R2GFormer, we have obtained the graph structural information and meta information for all nodes, including user, review and movie nodes. The next step is to fuse the multimodal information.

For each type of node, we utilize a type-wise TRM to facilitate inter-modal information interaction, then concatenate features of different modals followed by an MLP to get the final representation for each node, \textit{i.e.},
\begin{align}
    U_{g}, U_{m}, U_{b} = \text{TRM}([U_{g} \ U_{m} \ U_{b}]&, U = \text{\text{MLP}}([U_{g} \parallel U_{m} \parallel U_{b}])), \\
    R_{g}, R_{m} = \text{TRM}([R_{g} \ R_{m}]&, R = \text{\text{MLP}}([R_{g} \parallel R_{m}]) ), \\
    M_{g}, M_{m} = \text{TRM}([M_{g} \ M_{m}]&, M = \text{\text{MLP}}([M_{g} \parallel M_{m}]),
\end{align}
where $U_{g}$, $R_{g}$ and $M_{g}$ are derived from $N_{g}$, $U_{m}$, $R_{m}$ and $M_{m}$ are dervied from $N_{m}$.

After obtaining the comprehensive representation of each type of node, we then concatenate each review feature $r_{i}$ with its corresponding movie feature $m_{i}$ and user feature $u_{i}$:

\begin{equation}
    r_{i} = [\ r_{i} \parallel m_{i} \parallel u_{i} \ ].
\end{equation}

\subsection{GMoE}

Inspired by the successful applications of Mixture-of-Experts in NLP and bot detection, and its capability to handle the small subsets of the whole dataset, we adopt MoE to handle different genres of reviews. However, distinct from the latent subsets of the dataset in the classic MoE application scenario, our datasets already have the genre information of movies, as well as their related reviews. So we improve MoE to the proposed GMoE.

Specifically, instead of using the gating mechanism in the traditional MoE structure, we assign tokens to experts simply according to their genres: which genre it belongs to, which expert will deal with it; how many genres it belongs to, how many experts will deal with it.
\begin{align}
    r_{i} = \text{AGGREGATOR}_m(\{\text{Expert}_{j}(r_{i}) \mid \forall{j} \in{G_{i}}\}). 
\end{align}
where $G_{i}$ denotes the set of genres to which node $i$ belongs; AGGRE-\\GATOR\(_m\) can be summation, concatenation, or a TRM; each Expert is a MLP for simplicity.

\subsection{Learning and Optimization}

After using GMoE to process genre-specific information, we acquire the final representation $r_i$ for the $i$-th review. Then we apply a linear transformation to $r_i$ to obtain spoiler detection result $\hat{y_i}$. 
To train \ourmethod{}, We optimize the network by cross-entropy loss with $L_2$ regularization. The total loss function is as follows:

\begin{equation}
    Loss = -\sum_{i \in \mathcal{R}}{y_{i}\log \hat{y_{i}}} + \lambda\sum_{\theta \in \Theta}{\theta^{2}}.
\end{equation}
where $\hat{y_{i}}$ and $y_{i}$ are the prediction for the $i$-th review and its corresponding ground truth, respectively. $\mathcal{R}$ encompasses all the reviews in the training set, while $\Theta$ denotes all trainable model parameters in \ourmethod{}, and $\lambda$ is a hyper-parameter that maintains the balance between the two parts.

\section{Experiment}

\subsection{Experiment Settings}

\subsubsection{Dataset.}

To evaluate our \ourmethod{} framework along with 14 other representative baselines on two widely recognized datasets: \textbf{LCS}~\cite{wang2023detecting} and \textbf{Kaggle}~\cite{misra2019imdb}:

\begin{itemize}
    \item \textbf{LCS}  is a comprehensive dataset for automatic spoiler detection, comprising 1,860,715 reviews, 259,705 users, and 147,191 movies. And about 24.59\% (457,500) of the reviews are spoilers.
    \item \textbf{Kaggle}, introduced in 2019, consists of 573,913 valid reviews, 263,407 users, and 1,572 movies. And about 25.87\% (150,924) of the reviews are spoilers.
\end{itemize}

Note that both datasets include the genre information of all movies, with specific categories defined according to IMDb standards, which facilitates the operation of our \ourmethod{} framework. Following MVSD~\cite{wang2023detecting}, we randomly split the reviews into training, validation, and test sets with a ratio of 7:2:1.

\subsubsection{Baselines.}
To achieve a comprehensive evaluation, we compare \ourmethod{} with pre-trained language models, GNN-based models, and task-specific baselines. For the pre-trained language models, the procedure involves feeding the review text into the model, averaging all token embeddings, and then applying two fully connected layers to perform spoiler detection. Regarding the GNN-based models, the graph neural network takes the output of RoBERTa~\cite{liu2019roberta} as the initial node features. Below, we provide a concise overview of each baseline method.

\begin{itemize}
    \item \textbf{BERT}~\cite{devlin2018bert} is a language model pre-trained on extensive natural language data, using masked language modeling and next sentence prediction tasks.
    \item \textbf{RoBERTa}~\cite{liu2019roberta} improves upon BERT by eliminating the next sentence prediction task and enhancing masking techniques.
    \item \textbf{BART}~\cite{lewis2019bart} is a pre-trained language model that advances traditional autoregressive models through bidirectional encoding and denoising objectives.
    \item \textbf{DeBERTa}~\cite{he2021debertav3} refines BERT by implementing disentangled attention and an improved mask decoder, making it a more advanced language model.
    \item \textbf{Bge-Large}~\cite{xiao2023c} is trained on a comprehensive training dataset C-MTP, combining vast unlabeled data and diverse labeled data.
    \item \textbf{GCN}~\cite{kipf2016semi} is a foundational graph neural network that performs convolutions on graph nodes and their neighbors, effectively propagating information.
    \item \textbf{R-GCN}~\cite{schlichtkrull2018modeling} extends GCN to handle multi-relational graphs by incorporating relation-specific weights.
    \item \textbf{GAT}~\cite{velivckovic2017graph} is a graph neural network that applies attention mechanisms to dynamically assign importance to neighboring nodes.
    \item \textbf{SimpleHGN}~\cite{lv2021we} is tailored for heterogeneous graphs, integrating multiple types of nodes and edges with a shared embedding space and adaptive aggregation strategies.
    \item  \textbf{GPS}~\cite{rampavsek2022recipe} propose a recipe to build a general, powerful, scalable graph Transformer with linear complexity.
    \item \textbf{HGT}~\cite{hu2020heterogeneous} design node- and edge-type dependent parameters to characterize the heterogeneous attention over each edge for modeling Web-scale heterogeneous graphs.
    \item 
    \textbf{DNSD}~\cite{chang2018deep} is a spoiler detection method that employs a CNN-based genre-aware attention mechanism.
    \item \textbf{SpoilerNet}~\cite{wan2019fine} uses a hierarchical attention network and GRU alongside item and user bias terms for spoiler detection.
    \item \textbf{MVSD}~\cite{wang2023detecting} leverages external movie knowledge and user networks to detect spoilers.
\end{itemize}

\begin{table}[t]\scriptsize
\centering
\caption{Accuracy, AUC, and binary F1-score of \ourmethod{} and three types of baseline methods on two spoiler detection datasets. We run all experiments \textbf{five times} to ensure a consistent evaluation and report the average performance as well as standard deviation in parentheses. Bold indicates the best performance, \underline{underline} the second best. \ourmethod{} consistently outperforms the three types of methods on both benchmarks.}
\label{main-results}
\begin{tabular}{ccccccc}
\toprule
\textbf{Model} & \multicolumn{3}{c}{\textbf{Kaggle}} & \multicolumn{3}{c}{\textbf{LCS}} \\ \cline{2-7} 
 & F1 & AUC & Acc & F1 & AUC & Acc \\ \midrule
BERT & 44.02 $\pm$1.09 & 63.46 $\pm$0.46 & 77.78 $\pm$0.09 & 46.14 $\pm$2.84 & 64.82 $\pm$1.36 & 79.96 $\pm$0.38 \\ 
RoBERTa & 50.93 $\pm$0.76 & 66.94 $\pm$0.40 & 79.12 $\pm$0.10 & 47.72 $\pm$0.44 & 65.55 $\pm$0.22 & 80.16 $\pm$0.03 \\ 
BART & 46.89 $\pm$1.55 & 64.88 $\pm$0.71 & 78.47 $\pm$0.09 & 48.18 $\pm$1.22 & 65.79 $\pm$0.62 & 80.14 $\pm$0.07 \\ 
DeBERTa & 49.94 $\pm$1.13 & 66.42 $\pm$0.59 & 79.08 $\pm$0.09 & 47.38 $\pm$2.22 & 65.42 $\pm$1.08 & 80.13 $\pm$0.08 \\ 
Bge-Large & 52.51 $\pm$0.58 & 67.74 $\pm$0.37 & 77.44 $\pm$0.21 & 52.68 $\pm$0.36 & 68.46 $\pm$0.23 & 79.24 $\pm$0.11 \\ \midrule
GCN & 59.22 $\pm$1.18 & 71.61 $\pm$0.74 & 82.08 $\pm$0.26 & 62.12 $\pm$1.18 & 73.72 $\pm$0.89 & 83.92 $\pm$0.23 \\ 
R-GCN & 63.07 $\pm$0.81 & 74.09 $\pm$0.60 & 82.96 $\pm$0.16 & 62.99 $\pm$0.89 & 76.18 $\pm$0.72 & 85.19 $\pm$0.21 \\ 
GAT & 60.98 $\pm$0.09 & 72.72 $\pm$0.60 & 82.43 $\pm$0.01 & 65.73 $\pm$0.12 & 75.92 $\pm$0.13 & 85.18 $\pm$0.02 \\ 
SimpleHGN & 60.12 $\pm$1.04 & 71.60 $\pm$0.88 & 82.08 $\pm$0.26 & 63.79 $\pm$0.88 & 74.64 $\pm$0.64 & 84.66 $\pm$1.61 \\ 
HGT & 63.99 $\pm$0.25 & 75.61 $\pm$0.25 & 81.66 $\pm$0.23 & 60.89 $\pm$0.46 & 73.96 $\pm$0.53 & 81.86 $\pm$0.16 \\ 
GPS & 61.04 $\pm$0.84 & 73.50 $\pm$0.53 & 81.25 $\pm$0.55 & 64.21 $\pm$0.30 & 75.60 $\pm$0.92 & 82.40 $\pm$0.91 \\
\midrule
DNSD & 46.33 $\pm$2.37 & 64.50 $\pm$1.11 & 78.44 $\pm$0.14 & 44.69 $\pm$1.64 & 64.10 $\pm$0.74 & 79.76 $\pm$0.08 \\ 
SpoilerNet & 57.19 $\pm$0.69 & 70.64 $\pm$0.44 & 79.85 $\pm$0.10 & 62.86 $\pm$0.38 & 74.62 $\pm$0.69 & 83.23 $\pm$0.23 \\ 
MVSD & \underline{65.08} $\pm$0.69 & \underline{75.42} $\pm$0.56 & \underline{83.59} $\pm$0.11 & \underline{69.22} $\pm$0.61 & \underline{78.26} $\pm$0.63 & \underline{86.37} $\pm$0.08 \\ \midrule
\textbf{Ours}  & \textbf{80.24} $\pm$0.73 & \textbf{87.00} $\pm$0.37 & \textbf{89.65} $\pm$0.36 & \textbf{75.37} $\pm$0.10 & \textbf{83.71} $\pm$0.27 & \textbf{88.32} $\pm$0.08 \\ \bottomrule
\end{tabular}
\label{tab:performance}
\end{table}

\subsection{Main Results}

We evaluated our \ourmethod{} framework and 14 other baselines on two datasets. The results presented in Table \ref{main-results} demonstrate the following:
\begin{itemize}
    \item \textbf{\ourmethod{} consistently outperforms all baselines across both datasets}. Specifically, compared with the previous state-of-the-art method MVSD~\cite{wang2023detecting}, \ourmethod{} achieves \textbf{6.1\%} higher Binary-F1, \textbf{5.5\%} higher AUC, and \textbf{2.0\%} higher accuracy on the LCS dataset, as well as \textbf{15.2\%} higher Binary-F1, \textbf{11.6\%} higher AUC, and \textbf{6.1\%} higher accuracy on the Kaggle dataset. These improvements are statistically significant.
    \item In both datasets, \textbf{graph-based models generally outperform other types of baselines}, reaffirming the importance of analyzing the graph structure of reviews and their corresponding users and movies.
    \item Compared with DNSD~\cite{chang2018deep}, which also focuses on genre features of the reviews, \ourmethod{} surpasses DNSD across all three metrics in both datasets, further proving the effectiveness and robustness of our global-aware genreformer and GMoE methods.
    \item Both SpoilerNet~\cite{wan2019fine} and \ourmethod{} utilize user bias, but \ourmethod{} outperforms SpoilerNet in all three metrics across both datasets. This demonstrates that our dynamic graph pretraining can better identify the latent behavior pattern of whether a user is likely to post spoilers.
\end{itemize}

\subsection{Ablation Study}

As \ourmethod{} outperforms all the baselines and has reached state-of-the-art (SOTA) performance across the two datasets, we conducted ablation study to further explore the impact of each part of \ourmethod{} on the final performance with the Kaggle Dataset. The results are shown in Table \ref{ablation-results}.

\begin{itemize}
    \item To assess the importance of user bias information, we removed the user bias component $U_b$. The results in Table \ref{ablation-results} show an obvious decrease in performance, confirming that user bias information is critical for effective spoiler detection.  
    
    \item For the RetGAT component, we evaluated two variations: using an approximate method to compute $k$-hop neighbors~\cite{atwood2016diffusion,wang2020multi} and replacing our RetGAT with a standard GAT. Both variations lead to a drop in performance, indicating the necessity of our RetGAT design for capturing appropriate graph structures.
    
    \item To investigate the impact of GMoE, we replace it with a simple MLP, a traditional MoE~\cite{shazeer2017outrageously}, and a Soft-MoE~\cite{puigcerver2023sparse}. Note that we set the number of experts of the traditional  MoE and Soft-MoE to be the same as in GMoE, \textit{i.e.}, the number of genres. From the results shown in Table \ref{ablation-results}, the full model with GMoE achieves the best performance, highlighting the effectiveness of our GMoE design in handling explicit genre information. Moreover, replacing the GMoE with an MLP performs the worst, proving the rationality of using genre information.
    
    \item We also evaluate the Genreformer component by removing it and replacing the mean AGGREGATOR with sum pooling and max pooling. The results in Table \ref{ablation-results} show that the full Genreformer with sum AGGREGATOR outperforms these ablated versions, validating the necessity of the Genreformer.
    
\end{itemize}

\begin{table}[t]
    \centering
    \caption{Ablation study of \ourmethod{} on Kaggle Dataset. Bold indicates the best performance, \underline{underline} the second best.}
    \begin{tabular}{ccccc}
    \toprule
    \textbf{Catagory} & \textbf{Ablation Settings} & \textbf{F1} & \textbf{AUC} & \textbf{Acc} \\ 
    \midrule
    \textbf{User bias} & -w/o $U_b$ & 78.85 & 85.78 & 88.58 \\ 
    \midrule
    \multirow{2}{*}{\textbf{RetGAT}} 
    & approximate way & 78.40 & 85.39 & 88.39 \\ 
    & normal GAT & 78.08 & 85.22 & 88.29 \\ 
    \midrule
    \multirow{3}{*}{\textbf{GMoE}}
    & MLP & 78.52 & 86.09 & 88.25 \\
    & MoE~\cite{shazeer2017outrageously} & 78.75 & 86.36 & 88.54 \\
    & Soft MoE~\cite{puigcerver2023sparse} & 79.10 & 86.10 & 88.78 \\ 
    \midrule
    \multirow{3}{*}{\textbf{Genreformer}}
    & -w/o genreformer & 77.73 & 84.68 & 88.13 \\
    & sum pooling  & \underline{79.84} & \underline{86.56} & \underline{89.28} \\
    & max pooling  & 78.77 & 85.52 & 89.13 \\ \midrule
    \textbf{Ours} & \textbf{\ourmethod{}} & \textbf{80.24} & \textbf{87.00} & \textbf{89.65} \\
    \bottomrule
    \end{tabular}
    \label{ablation-results}
\end{table}

\subsection{GMoE Study} \label{Genre-Aware MoE Study}

In this section, we conduct further experiments on the GMoE to better understand its effectiveness compared to other previous MoE methods such as traditional MoE~\cite{shazeer2017outrageously} and Soft-MoE~\cite{puigcerver2023sparse}.

\subsubsection{Different numbers of Experts.}

\begin{figure}[tbp]
    \centering
    \includegraphics[width=\linewidth]{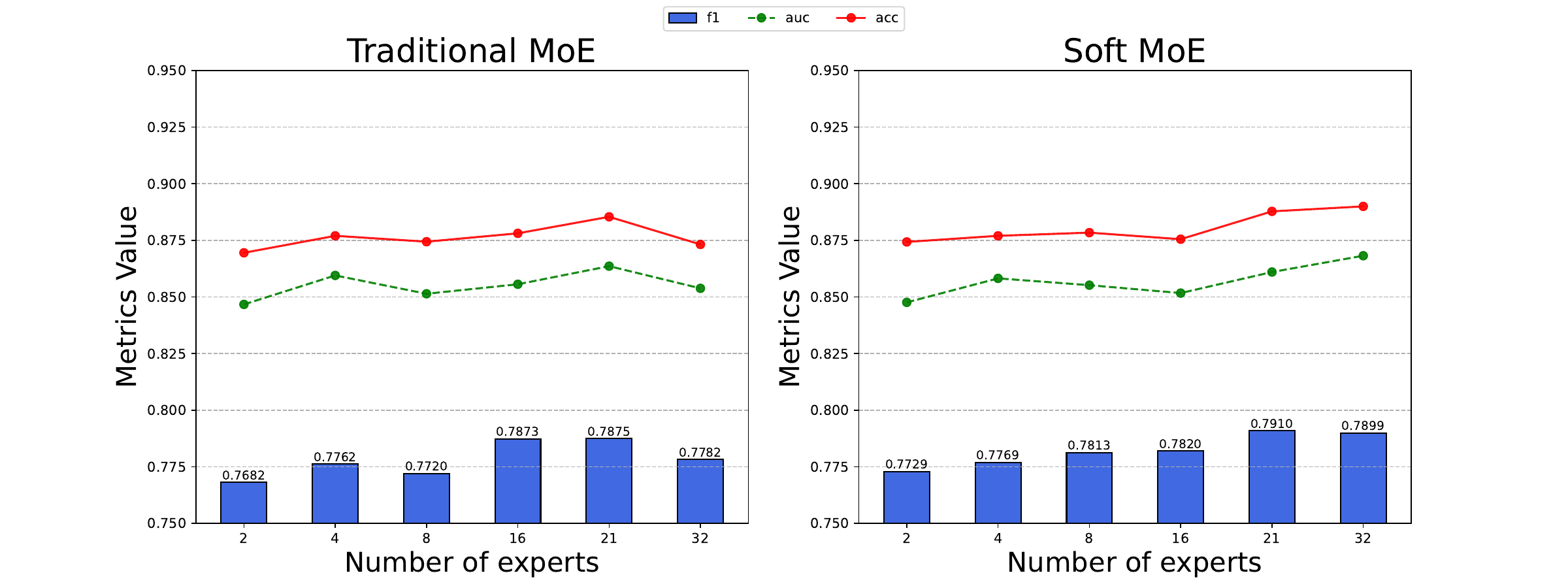}
    \caption{Performance comparison of different numbers of experts in traditional MoE and Soft-MoE. Note that 21 is the number of genres. The results indicate that GMoE outperforms other variants irrespective of the number of experts.}
    \label{fig:expert_comparison}
\end{figure}

We conduct multiple experiments by varying the number of experts in traditional MoE and Soft-MoE, recording the Binary-F1, AUC, and Accuracy for each configuration. The results are summarized in Figure \ref{fig:expert_comparison}.

From the results shown in Figure \ref{fig:expert_comparison}, it is evident that the GMoE consistently outperforms both traditional MoE and Soft-MoE regardless of the number of experts used. Specifically, Soft-MoE performs better than traditional MoE. That is because GMoE uses explicit genre information, eliminating inaccurate dispatch that can occur in traditional MoE and Soft-MoE.

\begin{figure}[t]
    \centering
    \includegraphics[width=0.7\linewidth]{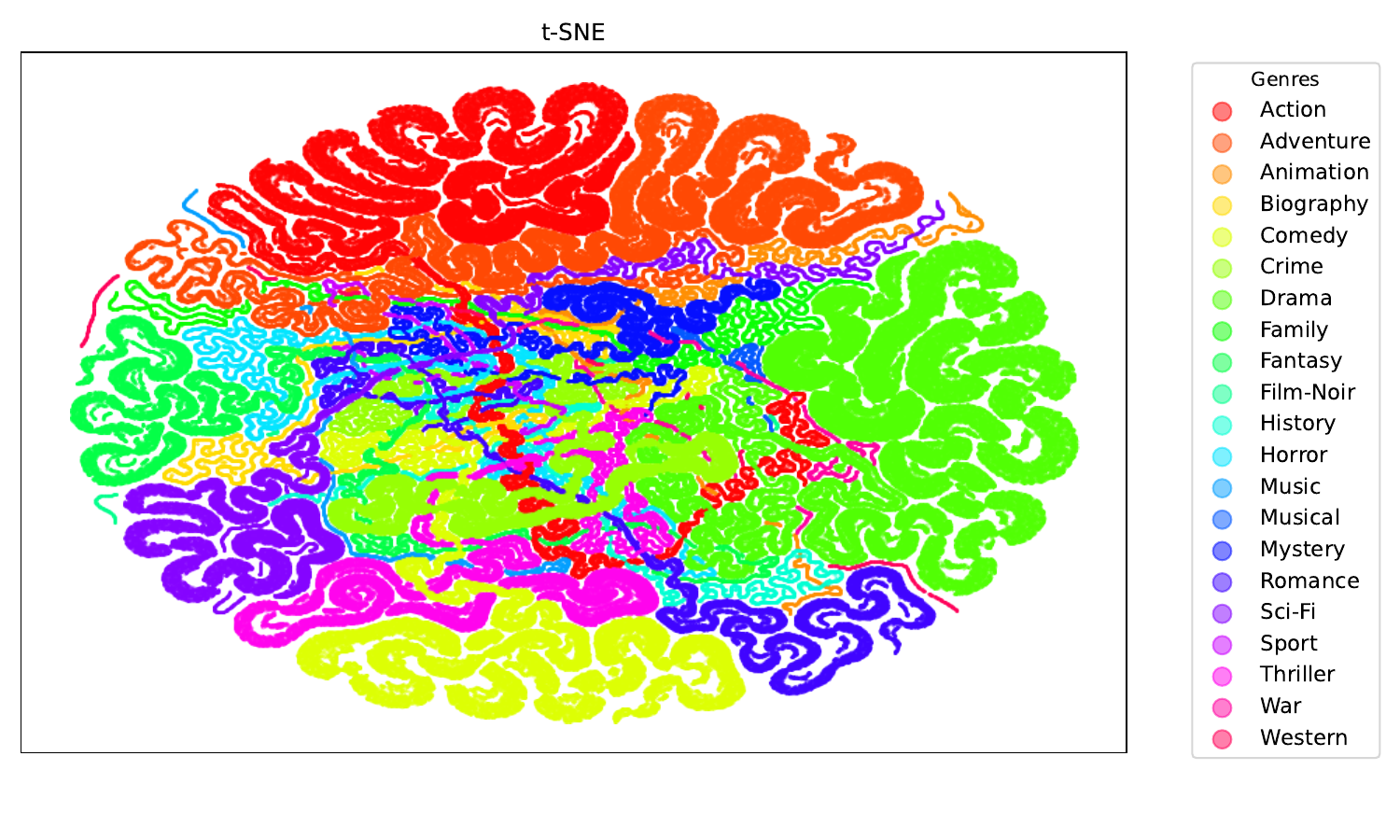}
    \caption{T-SNE visualization of the features processed by GMoE. Different colors represent different genres, indicating distinct clustering of features according to genres.}
    \label{tsne}
\end{figure}

\subsubsection{T-SNE Visualization of Features.}

To further validate the effectiveness of the GMoE, we perform dimensionality reduction on the features of all reviews processed by the GMoE layer. Specifically, we first reduce the dimensionality of the features to 50 dimensions using PCA, and then further reduce them to 2 dimensions using T-SNE. The result plot, shown in Figure \ref{tsne}, demonstrates that reviews of different genres exhibit distinct features after being processed by the GMoE.

The T-SNE visualization in Figure \ref{tsne} indicates that reviews are clustered according to their genres, providing compelling evidence of the effectiveness of our GMoE design. These distinct clusters suggest that GMoE successfully captures and utilizes genre-specific features to enhance its performance. In the visualization, each distinct color represents a different genre. For instance, genres like 'Action' (red), 'Drama' (green), and 'Sci-Fi' (purple) form well-defined clusters, indicating that the features of reviews from these genres are significantly different from each other. The presence of these distinct clusters reaffirms the model's capability to differentiate and leverage genre-specific information effectively.

\section{Conclusion}
In this paper, we introduce \ourmethod{}, a novel spoiler detection framework that integrates Genreformer and GMoE to effectively model diverse genre features. Additionally, \ourmethod{} incorporates dynamic graph pretraining to capture user bias related to spoiler posting. Extensive experiments reveal that \ourmethod{} significantly outperforms the state-of-the-art models on two major spoiler detection benchmarks. Further analysis validates the effectiveness of our proposed techniques, demonstrating \ourmethod{}'s superior capability in capturing intricate genre features and modeling user bias for spoiler detection.

\begin{credits}
\subsubsection{\ackname}

This work was supported by the National Nature Science Foundation of China (No. 62272374, No. 62192781 ), the Natural Science Foundation of Shaanxi Province (2024JC-JCQN-62), the National Nature Science Foundation of China (No. 62202367, No. 62250009), the Key Research and Development Project in Shaanxi Province No. 2023GXLH-024, Project of
China Knowledge Center for Engineering Science and Technology, and Project of Chinese academy of engineering “The Online and Offline Mixed Educational Service System for ‘The Belt and Road’ Training in MOOC
China”, and the K. C. Wong Education Foundation. Lastly, we would like to thank all LUD lab members for fostering a collaborative research environment.
\end{credits}

\newpage
\appendix

\section{Implementation Details} \label{Implementation Details}
We use PyTorch~\cite{paszke2019pytorch}, scikit-learn~\cite{pedregosa2011scikit}, PyTorch Geometric~\cite{fey2019fast}, MoE~\cite{rau2019moe}, soft-mixture-of-experts~\cite{puigcerver2023sparse} and 
Transformers ~\cite{wolf2020transformers} library to implement our proposed \ourmethod{}. We conduct our experiments on a cluster with 4 Tesla V100 GPUs with 32 GB memory, 16 CPU cores, and 377GB CPU memory. Code, data, hyperparameters and trained models will be made publicly available.

\section{More Experiments}

\begin{table}[h]
    \centering
    \caption{Evaluation of user bias effectiveness. The user bias pre-trained on Dynamic Graph achieves significantly higher accuracy in predicting user labels based on spoiler review percentage compared to raw user embeddings.}
    \begin{tabular}{cccc}
        \toprule
       \textbf{feature}  & \textbf{F1} & \textbf{AUC}  & \textbf{ACC} \\ \midrule
        DyG pretrained & 0.6389 & 0.7078 & 0.7777 \\ \midrule 
        raw & 0.5190 & 0.5532 & 0.5874 \\ \bottomrule
    \end{tabular}
    \label{User bias prediction}
\end{table}

\begin{table*}[htb]
    \centering
    \caption{Examples of the performance of top-2 baselines and \ourmethod{}. Underlined parts indicate the plots. "Key Information" indicates the most helpful information from other sources when detecting spoilers.}
    \begin{tabular}{m{4cm}<{\centering}m{3cm}<{\centering}cccc}
    \toprule
    \textbf{Review Text} & \textbf{Key Information} & \textbf{Label} & \textbf{R-GCN} & \textbf{MVSD} & \textbf{\ourmethod{}} \\ \midrule
    (...) The bomb is in a tower in both stories. Glen Manning and Bruce Banner are both there to watch the explosion. Glen and Bruce see some kid in the test area and both dash out to save him, tossing the boy into a ditch instead of jumping in with him. (...)   &  \makecell{\textbf{User Bias:} \\ Historical Reviews: \\ Review 1: spoiler \\ Review 2: spoiler} & \Checkmark & \XSolidBrush
& \XSolidBrush & \Checkmark \\ \midrule
    Don was an unsuccessful writer. Although he wanted to write a novel, he did not write any word. That should be mainly attributed to his alcoholism. It looked like that without drink he could not live.(...) & \makecell{\textbf{Movie Genres:}\\Drama\\Film-Noir} & \Checkmark & \XSolidBrush & \XSolidBrush & \Checkmark \\
    \bottomrule
    \end{tabular}
    \label{Case-Study}
\end{table*}

\begin{figure*}[ht]
    \centering
    \includegraphics[width=\linewidth]{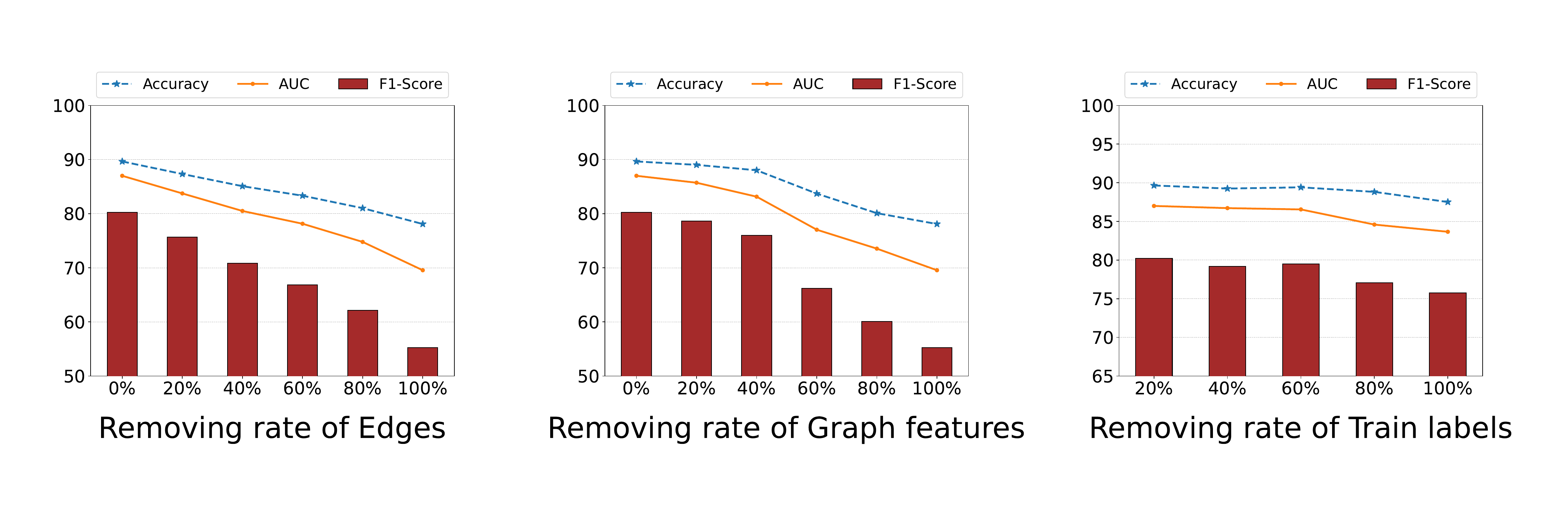} 
    \caption{When randomly removing edges from the graph, randomly setting elements of graph features to zero, and reducing the number of labeled samples in the training set, the performance of \ourmethod{} gradually declines with these incremental ablations. However, the fact that the model maintains a reasonable level of performance despite these perturbations indicates the robustness of our method.
}
    \label{fig:robustness_study}
\end{figure*}

\subsection{User Bias Analysis} \label{User Bias analysis}


To validate whether the obtained user bias truly reflects user preferences, we divide users into two categories based on whether more than 50\% of their reviews were spoilers. We then train a simple MLP using user bias as input to predict user labels. By contrast, we also use raw user embeddings that are not trained on a Dynamic Graph as input for predicting user labels. The results, shown in Table \ref{User bias prediction}, demonstrate that the predictions using pretrained user bias are significantly more accurate than those using the raw user embeddings. This indicates that user bias in \ourmethod{} effectively captures the users' preference tendencies.

\subsection{Robustness Study}
In the realm of social media analysis, particularly for tasks like spoiler detection, obtaining pure and clean data is a significant challenge. Spoiler detection often deals with information that is noisy and heterogeneous. As highlighted by previous studies on bot detection \cite{cai2024lmbot,liu2023botmoe}, it is crucial to investigate the robustness of models under conditions where information is incomplete or noisy.


To evaluate the robustness of \ourmethod{}, we introduce random perturbations to the input data to simulate scenarios where certain information might be missing. Specifically, for the graph structure, we randomly remove some edges; for the graph features, we randomly set some elements to zero; and for the training set, we randomly reduce the number of labeled samples. This experiment aims to assess the model's performance when parts of the multimodal information are incomplete.

Figure \ref{fig:robustness_study} presents the results of our experiments. As shown, even with some missing information, \ourmethod{} is still able to make correct predictions in most cases. This supports our hypothesis that multimodal information not only enhances the prediction accuracy of the model but also significantly improves its robustness.

\subsection{Case Study}

We conduct a case study to explore the impact of movie genres and user bias on spoiler detection. As illustrated in Table \ref{Case-Study}, our method \ourmethod{}, with the guidance of user bias and movie genres, correctly predicts spoilers, whereas the top-2 baseline models (R-GCN and MVSD) fail to do so.

Specifically, in the first case, our \ourmethod{}, bolstered by historical review data indicating consistent spoiler-posting behavior from the user, correctly identifies this review as a spoiler. In the second case, based on the Drama and Film-Noir movie genres, \ourmethod{} rightly determines that these elements reveal key plot points and hence labels it as a spoiler.

These examples underscore the importance of incorporating user bias and movie genre information. By effectively leveraging such data, \ourmethod{} demonstrates superior capability in identifying spoilers compared to baselines.

\begin{table}[ht]
    \centering
    \caption{Hyperparameter settings of \ourmethod{} in Kaggle}
    \begin{tabular}{lcc}
    \toprule
    \textbf{Hyperparameter} & \textbf{Kaggle} & \textbf{LCS} \\
    \midrule
    optimizer & AdamW  & AdamW \\
    learning rate & 1e-4 & 1e-4 \\
    weight decay & 1e-5 & 1e-5 \\
    CrossEntropyLoss weight & 1.3 & 1.3 \\
    epochs & 20 & 20 \\
    dropout & 0.3 & 0.3 \\
    batch size & 128 & 128 \\
    early stop patience & 3 & 3 \\
    \midrule
    LM  & Bge-Large \cite{xiao2023c}
 & Roberta \cite{liu2019roberta}\\    \midrule
    GNN layer & 2 & 3\\
    GNN input dim & 1024 & 768 \\
    GNN hidden dim & 1024 & 768 \\
    GNN output dim & 1024 & 768\\
    RetGAT hop num & 2 & 2 \\
    RetGAT hop decay $\alpha$ & 0.3 & 0.3 \\
    \midrule
    meta dim & 3 & 6 \\
    meta encoder hidden dim & 1024 & 768 \\
    meta encoder output dim & 1024 & 768 \\ 
    \midrule
    TRM heads & 4 & 4\\
    TRM layers & 2 & 2\\
    \midrule
    GMoE layer & 2 & 2\\
    GMoE hidden dim & 1024 & 768\\
    GMoE output dim & 1024 & 768\\
    GMoE experts number & 21 & 28 \\
    \bottomrule
    \end{tabular}
    \label{hyperparameter table}
\end{table}

\section{Hyperparameters}
We present our hyperparameter settings in Table \ref{hyperparameter table} to facilitate reproduction. Code, data,
and trained models will be made publicly available.

\begin{algorithm}[h]
\caption{$k$-hop Neighbors with Minimum Distance}
\label{$k$-hop Neighbors with Minimum Distance}
\begin{algorithmic}[1]
\State \textbf{Input:} adjacency matrix $A$, target $k$-hop number \textit{K}
\State \textbf{Output:} $k$-hop adjacency matrix $\mathcal{A}_k$

\State Initialize distance matrix $D$ with $\infty$, set diagonal to $0$
\State Initialize $\mathcal{A}_1$ with $A$
\ForAll{$(i, j)$}
    \If{$A_{ij} = 1$}
        \State $D_{ij} \gets 1$
    \EndIf
\EndFor
\For{$k \gets 2$ to \textit{K}}
    \State $A^k \gets A^{k-1} \times A$
    \ForAll{$(i, j)$}
        \If{$A^k_{ij} = 1$ \textbf{and} $D_{ij} > k$}
            \State $D_{ij} \gets k$
            \State $(\mathcal{A}_k)_{ij} \gets 1$
        \EndIf
    \EndFor
\EndFor

\State \Return $\mathcal{A}_k$

\end{algorithmic}
\end{algorithm}

\section{k-hop Neighbors Algorithm}

Previous studies \cite{atwood2016diffusion,wang2020multi} have used the $k$-th power of the adjacency matrix to capture $k$-hop visibility. This method does not strictly define $k$-hop neighbors as nodes reachable in exactly k steps. To address this, we introduce a new GPU-friendly \textbf{$k$-hop Algorithm} that quickly and accurately computes $k$-hop neighbors using adjacency matrices. The details are illustrated in Algorithm \ref{$k$-hop Neighbors with Minimum Distance}.

Considering that the number of $k$-hop neighbors $\mathcal{A}_k$ ($k>1$) can be large and may require substantial computational resources, we introduce an $r$-ratio (a hyper-parameter) to randomly sample neighbors from the entire set of neighbors for each hop.

\section{Metadata Collected for Nodes in Datasets}

Here we provide the details of metadata collected for nodes in two datasets in Table \ref{tab:metadata}

\begin{table}[h]
\centering
\caption{Comparison of metadata fields used in the LCS and IMDB datasets.}
\label{tab:metadata}
\begin{tabular}{llccc}
\toprule
\textbf{Entity} & \textbf{Field} & \textbf{Type} & \textbf{LCS} & \textbf{IMDB} \\
\midrule

\multirow{6}{*}{Movie} 
  & year       & int    & \checkmark & -- \\
  & isAdult    & bool   & \checkmark & -- \\
  & runtime    & int    & \checkmark & \checkmark \\
  & rating     & float  & \checkmark & \checkmark \\
  & numVotes   & int    & \checkmark & -- \\
  & genres     & list   & --         & \checkmark \\

\midrule

\multirow{7}{*}{User} 
  & badge\_count     & int    & \checkmark & -- \\
  & review\_count    & int    & \checkmark & \checkmark \\
  & avg\_helpful    & float  & \checkmark & -- \\
  & avg\_total      & float  & \checkmark & -- \\
  & avg\_score      & float  & \checkmark & -- \\
  & avg\_rating      & float  & --         & \checkmark \\
  & avg\_length      & float  & --         & \checkmark \\

\midrule

\multirow{5}{*}{Review} 
  & helpful\_votes        & int    & \checkmark & -- \\
  & total\_votes          & int    & \checkmark & -- \\
  & score                & float  & \checkmark & -- \\
  & rating               & float  & --         & \checkmark \\
  & review\_text\_length & int    & --         & \checkmark \\

\bottomrule
\end{tabular}
\end{table}

\end{document}